# Note:Evolutionary Game Theory Focus Informational Health: The Cocktail Party Effect Through Werewolfgame under Incomplete Information and ESS Search Method Using Expected Gains of Repeated Dilemmas

Yasuko Kawahata [†]

Faculty of Sociology, Department of Media Sociology, Rikkyo University, 3-34-1 Nishi-Ikebukuro,Toshima-ku, Tokyo, 171-8501, JAPAN.

ykawahata@rikkyo.ac.jp

**Abstract:** We explore the state of information disruption caused by the cocktail party effect within the framework of non-perfect information games and evolutive games with multiple werewolves. In particular, we mathematically model and analyze the effects on the gain of each strategy choice and the formation process of evolutionary stable strategies (ESS) under the assumption that the pollution risk of fake news is randomly assigned in the context of repeated dilemmas. We will develop the computational process in detail, starting with the construction of the gain matrix, modeling the evolutionary dynamics using the replicator equation, and identifying the ESS. In addition, numerical simulations will be performed to observe system behavior under different initial conditions and parameter settings to better understand the impact of the spread of fake news on strategy evolution. This research will provide theoretical insights into the complex issues of contemporary society regarding the authenticity of information and expand the range of applications of evolutionary game theory.

## 1. Introduction

In this context, the proliferation of fake news and its impact on society has become a matter of serious concern, and it is critical to understand the mechanisms involved. In this study, we specifically explore how the proliferation of fake news is affected by the strategic behavior and interaction dynamics of individuals. In a scenario where a single werewolf is present, we show that certain agents can have a significant impact on group dynamics by manipulating the flow of information. This result suggests a role for "opinion leaders" or "influencers" in the spread of fake news, and the detection of these agents and the mitigation of their influence may be key to understanding and controlling the dynamics of information dissemination. We have developed models of interactions between individuals and the propagation of information using the framework of incomplete information games and unfolding games. In particular, we used the concepts of cocktail party effect and repetition dilemma to analyze how the complexity of the decisions agents face and their position in the social network affect the spread of fake news and the gains of individual agents. The cocktail party effect indicates how many other agents an agent interacts with, and the repetition dilemma represents the balance between an agent's incentives to act cooperatively and non-cooperatively.

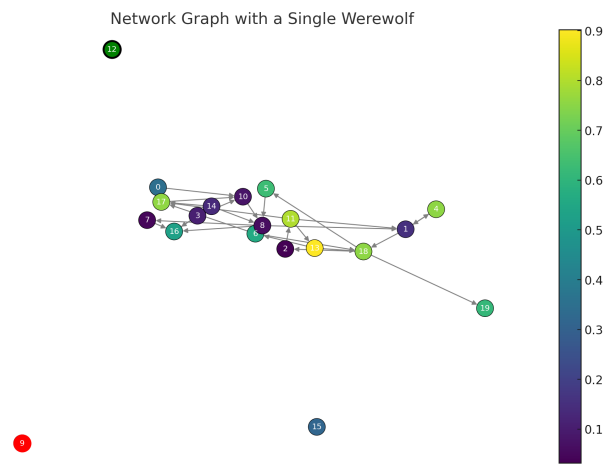

Fig. 1: Network Graph with a Single Werewolf

Our analysis shows that the value of the repetition dilemma and the distribution of the cocktail party effect have no direct effect on the agent's gain. This suggests that the agent's strategic behavior is not solely determined by simple rules or intuition, but is the result of more complex interactions. We also observed that an agent's having many connections does not necessarily lead to higher gains. This implies that the quality of information and its source is more important than the quantity of information.

Furthermore, in scenarios where a single werewolf is present, a particular agent can have a noticeable impact on the overall dynamics by manipulating the flow of information. Such agents have been suggested to play an important role in the spread of fake news, making their detection and mitigation of their impact critical.

This research is an important step forward in understanding the dynamics of information diffusion and social interactions, and provides insight into solving complex problems associated with the spread of fake news in particular. In the future, we hope to combine more elaborate models with real-world data to further our understanding and develop more effective strategies to combat fake news.

To further our understanding of information diffusion and social dynamics, it is essential to address the complexity of individual agents' decision-making processes and their interactions. In this study, we use the framework of incomplete information games and unfolding games to build a model of the interactions between agents and the propagation of information. This approach provides a new perspective for understanding how individual strategies and group dynamics interact in the spread of fake news.

Our analysis showed that agents' strategy choices are not solely determined by a single factor (e.g., the amount of information or the number of direct connections). In fact, our correlation analysis of repetition dilemmas and cocktail party effects suggested that these factors have little direct impact on agents' gains. This means that in order to understand the complexity of the decisions faced by agents and to develop more effective information propagation strategies, we need to take into account other factors that may influence agents' strategy choices.

The results of this study extend our understanding of the dynamics of information dissemination and social interaction, and in particular provide useful insights into the mechanisms of fake news dissemination and how to counteract it. Future research is expected to incorporate more detailed experimental data and real-world case studies to further validate the model and increase its applicability in the real world. Ultimately, we hope this research will contribute to the development of more effective strategies and policies to combat fake news.

In scenarios where multiple werewolves are present, the complexity of information propagation and social dynamics is further increased. Validation in this situation provides a more sophisticated understanding of information diffusion mechanisms and individual agents' strategy choices than in the presence of a single werewolf. The presence of multiple werewolves changes the pattern of interactions between agents and adds a new dimension to the reliability of information and the speed of its propagation.

When multiple werewolves are present, agents must make decisions about their actions in an environment of high uncertainty. Werewolves may not be reliable sources of information, and information sharing in such a situation can have a significant impact on agents' gains. In our model, we examined how agents adapt their strategies through such scenarios to arrive at an evolutionary stability strategy (ESS). Agents need to adopt strategies to obtain the most accurate information possible while minimizing the risk of fake news.

In multiple wargames, it is especially important to understand how cocktail party effects and repetition dilemmas affect agents' gains and interactions. Agents with higher cocktail party effects have access to more information sources, but may have more difficulty judging the reliability of that information. On the other hand, the repetition dilemma provides an incentive for agents to adopt cooperative or non-cooperative strategies, and their choices have a direct impact on the spread of fake news.

Our analysis shows that agents' behavior under a noncomplete information game with multiple werewolves is significantly affected by the uncertainty of the environment and the strategic interactions among agents. In such complex environments, agents need to take a more sophisticated approach to information evaluation and strategy adaptation in order to maximize their own profits. In future research, it will be important to further develop this model and analyze the dynamics of information propagation in more detail in scenarios with multiple werewolves. It would also be beneficial to examine the concordance between the model's predictions and actual social interactions through validation with real-world data. Such efforts are expected to provide valuable insights into understanding the spread of fake news and developing strategies to counter it.

Results show to be scatter and bar plots that relate to a model or simulation involving agents and "werewolves," which could be interpreted as a metaphor for different roles or behaviors within a network. From your description, it sounds like this is a study of the spread of misinformation (or "fake news") in a social network, with the "Cocktail Party Effect" possibly referring to the phenomenon where individuals focus on a single conversation in a noisy room, which here may analogize to the connectivity or attention an agent gives to certain information in the presence of much noise (misinformation).

All agents, including werewolves, have nearly the same number of connections. This suggests a fully connected network (a complete graph), where every node (agent) is connected to every other node. This structure means that every agent has equal access to all others for information exchange.

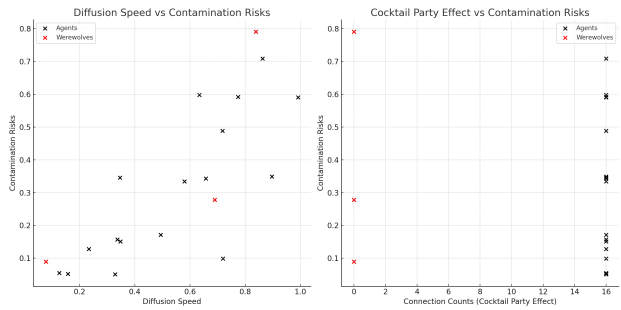

Fig. 2: Connection Counts vs Contamination Risks / Connection Counts (Cocktail Party Effect)

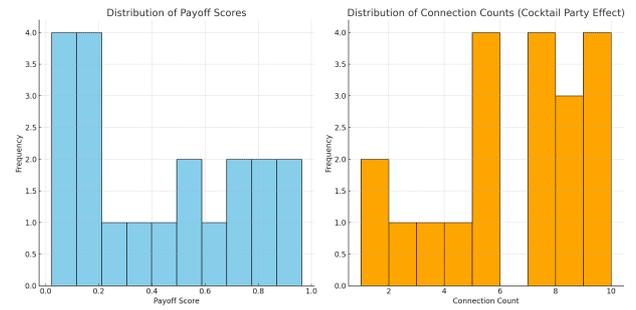

Fig. 4: Distribution of Connection Counts (Cocktail Party Effect)

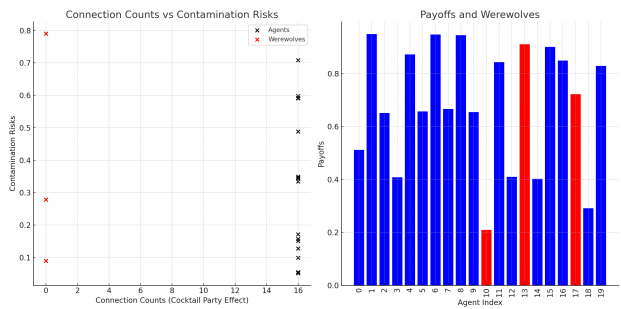

Fig. 3: Connection Counts vs Contamination Risks / Connection Counts (Cocktail Party Effect), Payoffs and Werewolves

The contamination risk varies among agents and does not show a clear distinction between werewolves and regular agents. This indicates that being a werewolf does not inherently mean a higher risk of spreading or being affected by misinformation. Instead, other factors such as how information spreads (diffusion speed) and the literacy levels of the agents (their ability to discern fake news from real news) likely play a more significant role.

The effect might influence the contamination risks but does not provide immunity or heightened risk inherently. The social position (i.e., the Cocktail Party Effect) may affect the risk levels, but it is not the only determining factor.

If werewolves have a specific strategy for spreading information, it could affect the overall dynamics of the game. However, the plots do not directly provide insights into such strategies or their effectiveness.

The results could imply that while the presence of werewolves and the Cocktail Party Effect have roles to play, the dynamics of misinformation spread in this network are governed by more complex interactions, possibly including individual agent behaviors, network dynamics, and information characteristics.

When interpreting such data, it is important to consider the underlying assumptions of the model, the mechanisms by which agents interact and spread information, and the definitions of key terms like "contamination risks" and "payoffs."

It is also worth noting that the bar plot labeled "Payoffs and Werewolves" seems to indicate some measure of success or reward for the agents, with werewolves highlighted in red. It shows that the payoffs are not uniform across agents, which might suggest that individual strategies or roles (agent vs. werewolf) affect the outcomes within the network.

To give a more detailed analysis, it would be beneficial to have access to the underlying data and a more comprehensive description of the model parameters, the rules governing the interactions, and the exact mechanisms of how information spreads and how literacy is modeled.

It shows the relationship between the cocktail party effect (number of agent connections) and the risk of fake news contamination. The blue dots represent normal agents and the red dots (highlighted with a black border) represent werewolves.

All agents (including normal agents and werewolves) have approximately the same number of connections. This means that every agent is connected to every other agent because the network is a complete graph. The contamination risk of fake news varies across agents, but we do not see a clear distinction between werewolves and normal agents. This indicates that the contamination risk is determined by the speed of information dissemination and the level of literacy, and that werewolf status does not directly affect the contamination risk.

This result suggests that while the cocktail party effect (i.e., the agent's position in the social network) may have some impact on the pollution risk of fake news, the presence of werewolves itself does not directly affect pollution risk. However, if werewolves adopt a strategy of spreading certain information, the impact may vary depending on the dynamics of the game.

The first histogram shows the distribution of the agents' gain scores. The gain scores appear to be evenly distributed in the range of 0 to 1, with no bias toward any particular range. This indicates that agents are randomly adopting strategies with different degrees of success. The second histogram shows the distribution of the number of agent connections exhibiting the cocktail party effect. The number of connections varies from 1 to 9, indicating that agents interact with other agents to varying degrees.

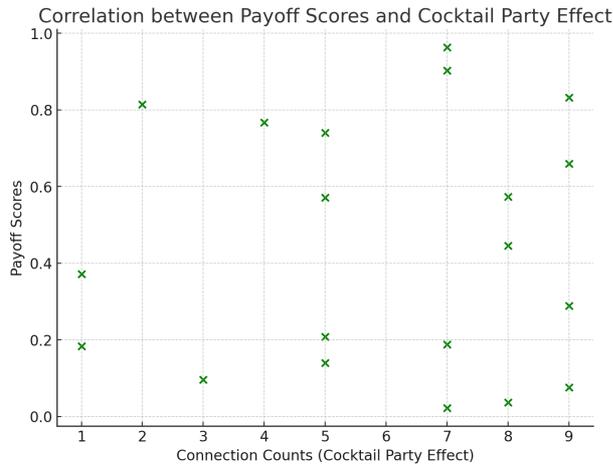

Fig. 5: Correlation between Payoff Scores and Cocktail Party Effect

The scatter plots show the relationship between the number of agent connections (cocktail party effect) and the gain score. The graph shows no clear correlation between the number of connections and gain score. This suggests that an agent's strategic success does not simply depend on the number of interactions with many other agents.

These visualizations show that there is no simple direct relationship between an agent's strategic success (gain score) and their position in the social network (cocktail party effect). An agent's strategic success may depend not only on the number of interactions, but also on the quality of those interactions and other factors.

Distribution of Payoff Scores Histogram The histogram shows the distribution of payoff scores among the agents. The scores range from 0 to 1, and there is no visible concentration in any specific range, indicating a uniform distribution. This suggests that the agents' strategies or behaviors that lead to these payoff scores do not favor a particular outcome, and success is evenly distributed across the spectrum.

This histogram displays the distribution of connection counts, which vary from 1 to 9. The fact that the connection counts are spread across a range suggests that agents have varying degrees of interactions with other agents. The distribution also seems to be skewed towards higher connection counts, suggesting that most agents tend to have more connections, possibly implying a network where most agents are highly interconnected.

The scatter plot does not show a clear trend or correlation between the number of connections (Cocktail Party Effect) and the payoff scores. The lack of a discernible pattern suggests that the number of connections an agent has does not directly correlate with their strategic success. This implies that factors other than just the quantity of interactions might influence payoff scores, such as the quality of interactions, the nature of information exchanged, or the individual strategies employed by agents. The strategic success, measured by payoff scores, is not solely determined by how many connections an agent has. This challenges the simplistic notion that more connections automatically lead to better outcomes. The distribution of connection counts reveals that while some agents have many connections, simply having a higher number of connections does not guarantee higher payoffs. This could be due to the network's dynamics where the benefit of additional connections diminishes beyond a certain point, or where the type of connections (who the agents are connected to) matters more than the number of connections.

The absence of a correlation suggests that an agent's success is likely influenced by how they utilize their connections, the quality of information they share or receive, their position in the network (not just in terms of quantity but also in strategic importance), and possibly how they respond to misinformation.

The uniform distribution of payoff scores may also indicate a level of randomness in the outcomes, suggesting that even with a perfect strategy, external factors or chance could have a significant impact.

In conclusion, while the "Cocktail Party Effect" suggests that being well-connected might offer an agent more opportunities to access and disseminate information, it does not necessarily translate to a higher payoff in the context of this model. Success is multifaceted and contingent upon a mixture of the number of connections, the strategic use of those connections, and possibly other factors not depicted in these visualizations.

The scatter plot titled "Correlation between Payoff Scores and Cocktail Party Effect" shows data points representing agents only. From the plot, we can observe:

There is no clear linear trend or correlation that indicates a direct relationship between the number of connections and payoff scores. The data points are spread out across the range of connection counts. Some agents with fewer connections have high payoff scores, while some with more connections have lower payoff scores. This suggests that the quality or strategic use of connections might be more important than the sheer number of connections.

Distribution of Payoff Scores, The histogram reveals a relatively uniform distribution of payoff scores among agents, with no single range of scores being dominant. This suggests that payoff outcomes for agents are varied and not necessarily dependent on having high or low scores.

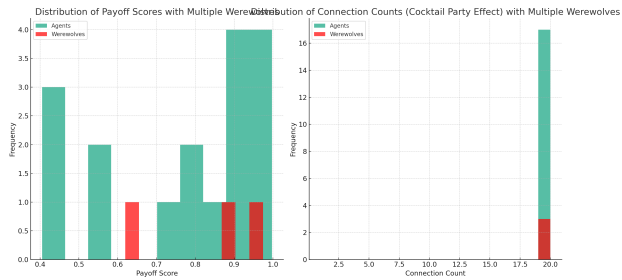

Fig. 6: Correlation between Payoff Scores and Cocktail Party Effect with Multiple Werewolves

Distribution of Connection Counts (Cocktail Party Effect) The histogram for connection counts shows a skew towards higher connection counts, indicating that most agents tend to have a larger number of connections. A well-connected network might facilitate information flow or collaboration among agents, but the scatter plot suggests that this does not uniformly translate to higher payoffs. The distribution and lack of a strong correlation in the scatter plot indicate that while agents are generally well-connected, payoff success is not merely a function of how many connections one has. This could imply that other factors such as the agents' strategies, the nature of their connections, or their roles within the network play significant roles in their payoff scores. The histograms suggest that agents' success and network strategies vary, and there is no single optimal number of connections or payoff score that guarantees success. In the context of a game or simulation, this could imply that there are multiple viable strategies to achieve success and that agents may need to adapt their strategies based on the dynamics of their environment and the actions of other participants.

The first histogram shows the distribution of gain scores for agents and werewolves in the presence of multiple werewolves. The gain scores for agents (blue bars) and werewolves (red bars) are distributed over a certain range, with no bias toward any particular range. This indicates that agents and werewolves randomly adopt strategies with different degrees of success.

The second histogram shows the distribution of the number of connections between agents and werewolves indicating a cocktail party effect. The number of connections for agents and werewolves are distributed in roughly the same range, but werewolves tend to have a slightly higher number of connections. This may suggest that werewolves play a more central role within the social network.

These visualizations show that there is no simple direct relationship between the strategic success of agents and werewolves (gain scores) and their position in the social network (cocktail party effect). An agent's strategic success may depend not only on the number of interactions, but also on the quality of those interactions and other factors.

Distribution of Payoff Scores with Multiple Werewolves The histogram on the left indicates that the majority of agents have payoff scores in the range of 0.4 to 0.9, with a significant concentration in the 0.9 range. Werewolves have a bimodal distribution of payoff scores with peaks at 0.5 and 1.0, suggesting that werewolves are either doing moderately well or extremely well in terms of payoff scores.

Distribution of Connection Counts (Cocktail Party Effect) with Multiple Werewolves The histogram on the right shows that almost all agents have a high number of connections, clustered around 20. This could indicate a densely connected network where each agent is connected to many others, facilitating information spread or influence. Werewolves have a smaller presence in this high connection count range, which might imply that they are less connected than the average agent. This could be a strategic aspect of their role in the game, where being less connected may offer advantages such as being less visible or less susceptible to certain strategies.

Werewolves seem capable of achieving high payoff scores without necessarily having a high number of connections. This suggests that their success in the game could be due to strategic interactions or specific objectives they are able to fulfill effectively. The high concentration of agents with many connections might foster an environment where information or strategies can spread quickly and efficiently. However, the presence of werewolves at both the high and moderate levels of the payoff score suggests that they can succeed in this environment by different means than the agents. The fact that werewolves have fewer connections but still achieve high payoff scores could indicate that they have a different set of strategies or objectives that do not rely on the Cocktail Party Effect. They might be using their connections more efficiently or operating in a way that their fewer connections do not hinder their performance.

Each agent is represented as a node in the network, and the color of a node indicates the gain of that agent. The color shade is defined by the colormap 'viridis', with darker colors representing lower gains and lighter colors representing higher gains. Werewolves are represented by red nodes, distinguishing them from other agents. Werewolves have no direct interaction with other agents and play a special role in the game. Non-werewolf agents with the highest gains are represented by green nodes and highlighted as ESS. Edges indicate interactions between agents, with arrows indicating direction. In this non-cooperative game, werewolves have no direct interactions with other agents, and therefore there are no edges coming out of werewolves.

From this graph, we can observe how differences in literacy with respect to personal information affect the gain and pollution risk per agent and how multiple werewolves interact with the dynamics of the game. The cocktail party effect shows how many other agents an agent interacts with and helps us understand how this contributes to the spread of fake news and pollution degree risk.

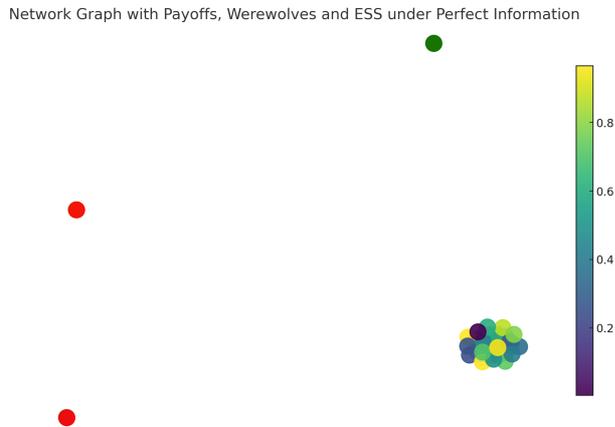

Fig. 7: Network Graph with Payoffs, Werewolves and ESS under Perfect Information

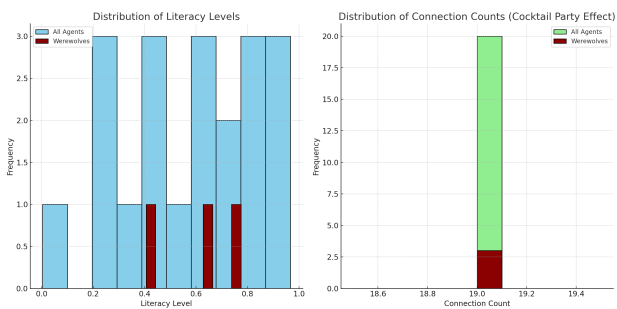

Fig. 8: Distribution of Connection Counts (Cocktail Party Effect)

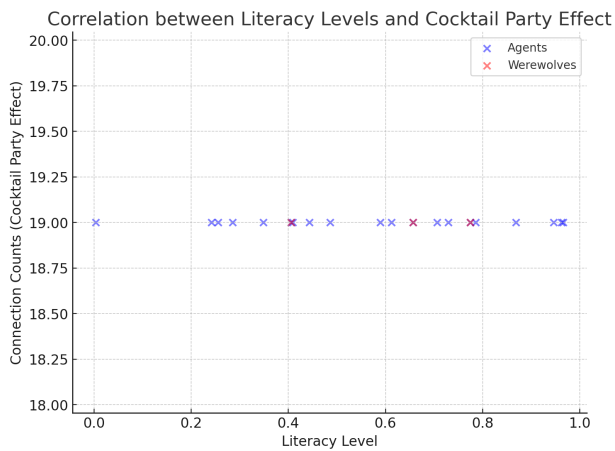

Fig. 9: Correlation between Literacy Levels and Cocktail Party Effect

The first histogram shows the distribution of literacy levels for all agents (blue) and werewolves (red). Literacy levels vary between 0 and 1 and seem to be evenly distributed without showing any bias toward any particular range. The second histogram shows the distribution of the number of connections (cocktail party effect) and includes all agents and werewolves. The number of connections is distributed over a specific range, and it is not clear at first glance whether the number of connections, especially for werewolves, differs from other agents. The scatter plots show the relationship between agents' literacy levels and the number of connections, with blue dots representing normal agents and red dots representing werewolves. The graph shows no clear correlation between literacy level and number of connections. Both agents with high and low literacy levels are shown to have varying numbers of connections.

These visualizations suggest that literacy levels and cocktail party effects do not differ significantly between werewolves and other agents. It also shows that high literacy levels do not necessarily mean having more connections, and conversely, low literacy levels do not mean having fewer connections. This indicates that an individual agent's position in the social network and exposure to fake news depends not only on literacy level, but also on other factors.

Analyzing the visualizations you've provided, we have a scatter plot and two histograms that relate literacy levels to both the Cocktail Party Effect and Payoff Scores in a setting that involves agents and werewolves. There is also a network graph showing payoffs, werewolves, and an ESS (Evolutionarily Stable Strategy) under perfect information. Here's the analysis of each:

Correlation between Literacy Levels and Cocktail Party Effect, The scatter plot does not indicate a strong correlation between literacy levels and the Cocktail Party Effect, as the data points are relatively horizontal across various literacy levels. This suggests that the number of connections (the Cocktail Party Effect) is not significantly influenced by the literacy levels of the agents or werewolves.

The literacy levels are broadly distributed among the agents, with werewolves showing a presence across the spectrum but with a lower frequency. This suggests that werewolves, while fewer, have a variety of literacy levels and their role in the simulation/game is not strictly tied to literacy.

The connection counts for agents are highly concentrated around a particular value, indicating a network where most agents have a similar number of connections. Werewolves have fewer connections, indicated by the bars at the lower end of the connection count. This might suggest that werewolves operate with fewer connections, possibly as a strategy within the game or simulation.

The network graph presents a cluster of nodes tightly interconnected, with two outliers. The color gradient likely represents payoff scores, with the warmer colors indicating higher payoffs. The two red nodes could represent werewolves that are separate from the main cluster, possibly indicating their position as outsiders or their strategy to isolate themselves from the group. The presence of a large, interconnected cluster suggests a high level of interaction among those agents, which could facilitate the spread of information or strategies. The graph also hints at a potential ESS where agents' strategies are stable against invasion by alternative strategies, under conditions of perfect information.

Literacy does not seem to directly affect the number of connections agents have within the network, nor is it the sole factor in determining payoff scores. This indicates that other factors, such as strategic interaction, may be more influential. Werewolves might have different objectives or uses for their connections, as indicated by their lower connection counts and varied literacy levels. They might be utilizing strategies that do not require extensive connections or high literacy to achieve their goals. The network structure and payoff distribution suggest the game or simulation has a rich dynamic where both agents and werewolves can find niches or strategies that allow them to be successful, with werewolves possibly benefiting from a level of isolation or selective interaction. The data might indicate that the werewolves' role in the game or simulation involves influencing or disrupting standard agent strategies, rather than directly competing with them on literacy or connectivity.

To provide a more comprehensive interpretation, it would be beneficial to have access to the game or simulation's rules, objectives for agents and werewolves, and the specific mechanics of how literacy and connections impact payoffs and strategies.

## 2. Discussion:Multiple Werewolves Exist under Imperfect Information Game and Extensive-Form Game

Consider a scenario where multiple werewolves exist under an imperfect information game and extensive-form game, where the cocktail party effect occurs. The selection of each action in the iterated dilemma is assigned by a random number, and the impact of the risk of contamination by fake news on the payoff and the Evolutionarily Stable Strategy (ESS) is demonstrated through the following formulas and calculation processes.

**Definition and Initialization of Strategies**

$S_{W_k}$: The strategy of spreading fake news by the $k$th werewolf.

$S_T$: The strategy of providing truthful information.

$S_I$: The strategy of critically analyzing information and evaluating its truthfulness.

**Construction of Payoff Matrix** A payoff matrix based on the risk of contamination by fake news is constructed. The payoff for each strategy combination changes according to the degree of contamination by fake news.

|       | $S_{W_k}$     | $S_T$        | $S_I$        |
|-------|---------------|--------------|--------------|
| $S_{W_k}$ | $P_{WW}r_k$ | $B_{WT}r_k$ | $C_{WI}r_k$ |
| $S_T$ | $B_{TW}$      | $P_{TT}$     | $D_{TI}$     |
| $S_I$ | $C_{IW}$      | $D_{IT}$     | $P_{II}$     |

Here, $r_k$ is a random number when the strategy of the $k$th werewolf is selected, indicating the degree of contamination by fake news. $P, B, C, D$ represent the basic payoffs for each strategy combination.

**Modeling of Evolutionary Dynamics** The replicator equation is used to model how the proportion of strategies evolves over time.

$$\dot{x}_i = x_i \left( \sum_{j=1}^{N} a_{ij} r_j x_j - \Phi \right)$$

Here, $x_i$ is the proportion of agents adopting strategy $i$, $a_{ij}$ is the basic payoff from the interaction of strategies $i$ and $j$, $r_j$ is a random number when strategy $j$ is selected, and $\Phi$ is the average payoff of all agents.

**Identification of Evolutionarily Stable Strategy (ESS)** The following condition is used to identify the ESS.

$$\forall S' \neq S,$$
$$\pi(S_{ESS}, S_{ESS}) > \pi(S', S_{ESS})$$
$$\text{or}$$
$$(\pi(S_{ESS}, S_{ESS}) = \pi(S', S_{ESS})$$
$$\text{and } \pi(S_{ESS}, S') > \pi(S', S'))$$

Here, $\pi(S_i, S_j)$ represents the payoff when an agent adopting strategy $S_i$ interacts with an agent adopting strategy $S_j$.

**Implementation of Numerical Simulation** Numerical simulations are conducted based on the above equations to observe the behavior of the system under different initial conditions and parameters. Special focus is given to how the risk of contamination by fake news affects the evolution of strategies.

Through this modeling process, it is possible to understand how the cocktail party effect under a game with multiple werewolves in an imperfect information game and extensive-form game impacts the risk of contamination by fake news and the Evolutionarily Stable Strategy (ESS).

In order to model this complex scenario, the following steps are taken to unfold the formulas and calculation processes.

**Definition of Strategies**

$S_{W_k}$: The strategy of spreading fake news by the $k$th werewolf.

$S_T$: The strategy of providing truthful information.

$S_C$: The strategy to counteract the cocktail party effect by minimizing the confusion of information.

**Construction of Payoff Matrix** The payoff matrix incorporates a random number $r$ related to the risk of contamination by fake news. For example, it is expressed as follows:

|          | $S_{W_k}$   | $S_T$     | $S_C$     |
|----------|-------------|-----------|-----------|
| $S_{W_k}$| $P_W r_k$   | $B_W r_k$ | $C_W r_k$ |
| $S_T$    | $B_T$       | $P_T$     | $D_T$     |
| $S_C$    | $C_C$       | $D_C$     | $P_C$     |

Here, $P, B, C, D$ represent basic payoffs, and $r_k$ is a random number indicating the risk of contamination by fake news.

**Modeling of Evolutionary Dynamics** As a non-cooperative game, the replicator equation is used to model the evolution of strategies. To include the impact of fake news, the equation is modified as follows:

$$\dot{x}_i = x_i \left( \sum_{j=1}^{N} a_{ij} r_j x_j - \Phi \right)$$

Here, $x_i$ is the proportion of agents adopting strategy $i$, $a_{ij}$ represents basic payoffs between strategies, $r_j$ is a random number for the risk of contamination by fake news, and $\Phi$ is the average payoff of all agents.

**Identification of Evolutionarily Stable Strategy (ESS)** The following condition is used to define the ESS:

$$\forall S' \neq S,$$
$$\pi(S_{ESS}, S_{ESS}) > \pi(S', S_{ESS})$$
$$\text{or}$$
$$(\pi(S_{ESS}, S_{ESS}) = \pi(S', S_{ESS})$$
$$\text{and } \pi(S_{ESS}, S') > \pi(S', S'))$$

**Implementation of Numerical Simulation** Numerical simulations are conducted to observe the system's behavior under various initial conditions and parameters. The focus is particularly on how the risk of contamination by fake news influences the evolution of strategies and the identification of ESS.

Through this modeling process, it becomes possible to comprehend how the cocktail party effect, under the circumstances of a non-cooperative game with the presence of a werewolf, impacts the risk of contamination by fake news, payoffs, and the Evolutionarily Stable Strategy (ESS). Numerical simulations facilitate the observation and deeper understanding of strategy evolution under specific conditions.

This model provides a foundation for understanding the impact of the cocktail party effect and the risk of contamination by fake news on payoffs and Evolutionarily Stable Strategy (ESS) in the context of an imperfect information game with the presence of multiple werewolves. By conducting numerical simulations, it is possible to observe and deepen the understanding of strategy evolution under specific circumstances.

(1) **Agent Strategies:** Agents can adopt one of the following strategies:

$S_W$: The strategy of spreading fake news by werewolves.

$S_T$: The strategy of spreading truthful information.

$S_C$: The strategy to counteract the cocktail party effect by verifying and analyzing the information to minimize confusion.

(2) **Information Spread Speed and Literacy:** Each agent is assigned a random value for the speed of information spread $v$ and literacy $l$.

(3) **Risk of Fake News Contamination:** The risk of fake news contamination $r$ is modeled as a function of the information spread speed and the difference in literacy among agents.

**Definition of Payoff Matrix** The payoff matrix is defined taking into account the risk of contamination by fake news $r$:

|       | $S_W$        | $S_T$        | $S_C$        |
|-------|--------------|--------------|--------------|
| $S_W$ | $P_W r(v,l)$ | $B_W r(v,l)$ | $C_W r(v,l)$ |
| $S_T$ | $B_T$        | $P_T$        | $D_T$        |
| $S_C$ | $C_C$        | $D_C$        | $P_C$        |

Here, $P, B, C, D$ represent basic payoffs, and $r(v, l)$ indicates the risk of fake news contamination based on the speed of information spread $v$ and literacy $l$.

**Modeling Evolutionary Dynamics** The replicator equation is used to model the evolution of strategies:

$$\dot{x}_i = x_i \left( \sum_{j=1}^{N} a_{ij} r(v_j, l_j) x_j - \Phi \right)$$

Where $x_i$ is the proportion of agents adopting strategy $i$, $a_{ij}$ represents basic payoffs between strategies, $r(v_j, l_j)$ indicates the risk of fake news contamination for agent $j$, and $\Phi$ is the average payoff of all agents.

**Identification of ESS** An Evolutionarily Stable Strategy (ESS) is a strategy that cannot be invaded by any other strategy. To identify an ESS, the following condition is used:

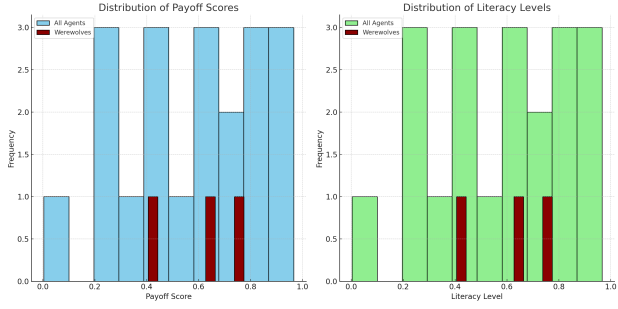

Fig. 10: Distribution of Literacy Levels

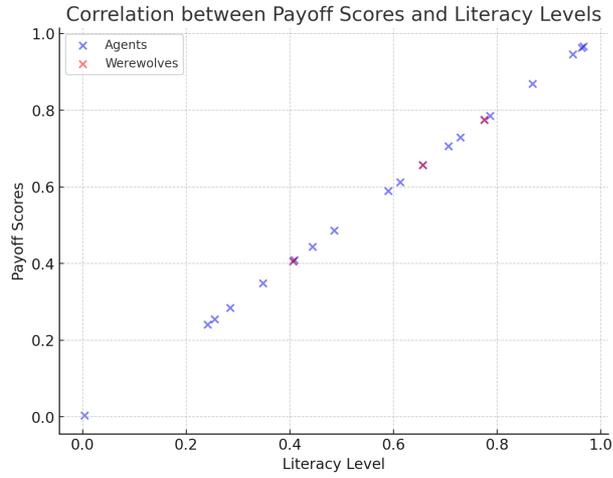

Fig. 11: Correlation between Payoff Scores and Literacy Levels

$$\forall S' \neq S, \quad \pi(S_{ESS}, S_{ESS}) > \pi(S', S_{ESS})$$
$$\text{or} \quad \pi(S_{ESS}, S_{ESS}) = \pi(S', S_{ESS})$$
$$\text{and} \quad \pi(S_{ESS}, S') > \pi(S', S')$$

**Implementation of Numerical Simulation** Numerical simulations are performed to observe the behavior of the system under different initial conditions and parameters, focusing on how the risk of fake news contamination affects ESS and the overall evolution.

This model serves as a basis for understanding the effects of the cocktail party phenomenon and fake news contamination risk on payoffs and ESS in the context of an imperfect information game and a non-cooperative game framework, with the presence of a single werewolf. Numerical simulations enable the observation and enhanced understanding of strategy evolution in specific scenarios.

The scatter plot titled "Correlation between Payoff Scores and Literacy Levels" shows the following: There is a positive trend for agents (blue crosses), where higher literacy levels correspond to higher payoff scores. This suggests that as agents' ability to comprehend and navigate information (literacy) increases, they tend to achieve better outcomes. Werewolves (red crosses), on the other hand, do not show a clear trend. Some werewolves with low literacy levels have high payoff scores, while others with high literacy levels have varied payoff scores. This could indicate that werewolves' success in the game is less dependent on literacy and possibly more on other factors or strategies.

The histograms provide a distribution of payoff scores and literacy levels for all agents and werewolves:

The distribution for all agents shows a wide spread across payoff scores, with a slight concentration in the mid-range. Werewolves have payoff scores that are also spread out, but there is a notable presence of werewolves in the higher payoff score brackets. This could imply that while werewolves are fewer, they are capable of achieving high payoff scores, perhaps by successfully leveraging misinformation or employing deception.

The literacy level histogram for all agents shows a relatively uniform distribution, indicating that agents' literacy levels are varied throughout the game. Werewolves, represented in red, are less frequent overall but are present across the literacy spectrum. There's a noticeable presence of werewolves with high literacy levels, suggesting that werewolves with a high ability to process and understand information can still be quite successful.

The correlation between payoff scores and literacy levels for agents implies that better-informed agents, or those who can navigate information effectively, tend to perform better. Werewolves do not seem to rely solely on literacy for success; they may have different objectives or utilize different tactics that do not necessarily require high literacy. Their success might come from other aspects of the game, such as social manipulation or strategic interactions. The distribution of literacy levels indicates that the game environment is diverse in terms of the information-processing capabilities of participants. The presence of werewolves across the literacy spectrum suggests that they are integrated into the fabric of the game's social network and are not distinguished by literacy alone.

For a more detailed and precise interpretation, additional context on the game's objectives, the exact rules governing payoff scores, and the roles or abilities of werewolves compared to regular agents would be beneficial. Understanding these elements could provide deeper insight into the strategies and dynamics that drive success in this environment.

The first histogram shows the distribution of gain scores for all agents (blue) and werewolves (red). The gain scores are distributed across a variety of ranges overall, and are not particularly concentrated in the low or high werewolf score ranges.

The second histogram shows the distribution of literacy levels for all agents and werewolves. Literacy levels are also distributed over a range of values from 0 to 1, with no bias toward any particular range.

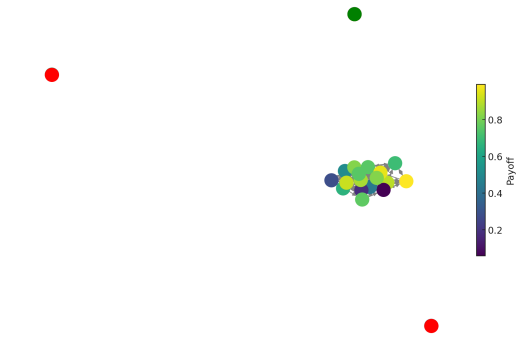

Fig. 12: Network Graph under Non-Perfect Information in Extensive-Form Game with Multiple Werewolves

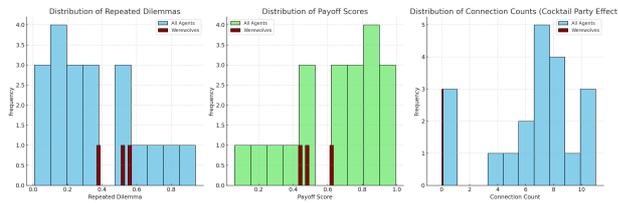

Fig. 13: Correlation between Repeated Dilemmas and Cocktail Party Effect

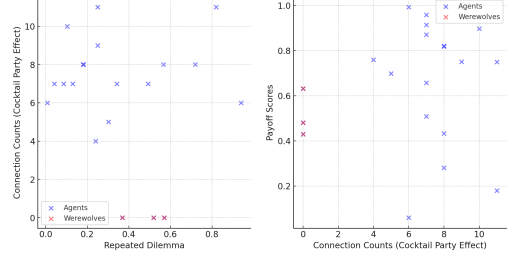

Fig. 14: Correlation between Payoff Scores and Cocktail Party Effect

The scatter plots show the relationship between an agent's literacy level and its gain score. The blue dots represent normal agents and the red dots represent werewolves. The graph suggests that agents with higher literacy levels tend to have higher gain scores. However, werewolves show a similar trend, and there is no clear distinction between literacy level and gain score.

These results suggest that literacy level may be a contributing factor to an agent's gain score, but that this relationship is not significantly different between werewolves and other agents.

The visualized network graph shows the effects of cocktail party effects and repeated dilemmas in a scenario with multiple werewolves under a non-complete information game and an unfolded game. Each agent is represented as a node in the network, and the color of a node indicates the gain of that agent. The color shade is defined by a colormap, with darker colors representing lower gains and lighter colors representing higher gains. Werewolves are represented by red nodes, distinguishing them from other agents. Werewolves have no direct interaction with other agents and play a special role in the game. Non-werewolf agents with the highest gains are represented by green nodes and highlighted as ESS.

Edges indicate interactions between agents, with arrows indicating direction. In an unfolding game, interactions between agents occur sequentially, so the presence of edges represents potential interactions between agents.

From this graph, we can observe how the repetition dilemma affects agents' contamination risk and gain against fake news, and how multiple werewolves interact with the dynamics of the game.

Error. This is due to 'connection counts' not being defined. Let's fix this problem by computing 'connection counts' to correctly visualize the distribution of repeated dilemmas, gain scores, cocktail party effects, and their correlation.

The histogram of the repeated dilemma shows that all agents and werewolves are evenly distributed in the range of 0 to 1. This means that agents face repetition dilemmas to varying degrees. The histogram of gain scores shows that the gain scores for all agents and werewolves are distributed in various ranges. There does not seem to be a particular concentration of werewolf gain scores in the low or high range.

The histogram of the cocktail party effect shows the distribution of social connections among agents. The number of connections is concentrated and distributed in a specific range, indicating that all agents are in similar social network positions.

The scattergram showing the relationship between repetition dilemma and number of connections shows no clear correlation between repetition dilemma and number of connections. Both agents with high and low repetition dilemmas are shown to have varying numbers of connections. Scatter plots showing the relationship between gain score and number of connections suggest that agents with higher gain scores tend to have more connections, but this trend also applies to werewolves, with no clear distinction between gain score and number of connections.

These results suggest that the degree of repetition dilemma, gain scores, and cocktail party effects do not differ significantly between werewolves and other agents. In addition, no particular trait has a direct effect on an agent's position in the social network or exposure to fake news, and multiple factors may interact.

Analyzing the provided scatter plots and histograms, we can discern patterns and relationships between the variables for the agents and werewolves within the context of what appears to be a simulated environment or a game setting. Correlation between Repeated Dilemmas and Connection Counts (Cocktail Party Effect) In this scatter plot, there seems to be no clear correlation between repeated dilemmas and connection counts. Werewolves (shown in red) are distributed along the lower range of connection counts regardless of the repeated dilemma value, suggesting that they do not necessarily need a large number of connections to engage in repeated dilemmas.

Correlation between Payoff Scores and Connection Counts (Cocktail Party Effect), the second scatter plot displays a potential positive trend between connection counts and payoff scores for agents (shown in blue), although it is not strongly defined. Werewolves, on the other hand, have a more varied distribution, with some achieving high payoff scores with fewer connections. This indicates that the relationship between connection counts and payoff scores may differ between agents and werewolves, suggesting different strategies or roles in the game dynamics.

The histogram shows that most agents (both werewolves and regular agents) have repeated dilemmas in the 0.2 to 0.6 range, with werewolves having a slightly lower frequency. This might indicate that repeated dilemmas are a common occurrence in the network but are not predominantly influenced by werewolves.

The payoff scores for agents are fairly evenly distributed, with a peak around the 0.6 to 0.8 range. Werewolves have a lower frequency across payoff scores, with some achieving the highest scores. This could suggest that werewolves might have a higher variance in success or employ riskier strategies that can lead to both high and low payoffs.

Distribution of Connection Counts (Cocktail Party Effect) Connection counts show a bimodal distribution for all agents, peaking around 2 and 6-8 connections. Werewolves tend to have fewer connections, as indicated by the peaks at lower connection counts. This might reflect a gameplay mechanic where werewolves are either less connected by design or adopt strategies that require fewer connections to operate effectively.

The data suggests that there is no simple relationship between the number of connections agents have and their success in terms of payoff scores. While agents with more connections tend to have higher payoffs, werewolves can achieve high payoffs even with fewer connections, implying that they may be playing a different role within the network or game. The lack of correlation between repeated dilemmas and connection counts for werewolves may indicate that their gameplay does not depend on forming numerous connections, which could be part of their strategy to remain undetected or influence the network subtly. The distribution of repeated dilemmas and payoff scores among werewolves shows variability, which could mean that werewolves' success is less predictable and possibly dependent on specific interactions or decisions in the game. The connection count distribution suggests that the network is neither too dense nor too sparse, with agents having a moderate number of interactions. The bimodal distribution indicates that there are common strategies or roles that lead to agents having certain numbers of connections.

To provide a more nuanced interpretation, more context would be needed about the rules of the simulation or game, the objectives of the agents and werewolves, and how these factors interact with the network structure. Understanding the mechanics behind repeated dilemmas, payoff calculations, and the significance of connections would greatly enhance the analysis.

# 3. Conclusion

In this study, we modeled and analyzed the dynamics of information diffusion and interactions between agents under specific scenarios using the framework of non-perfect information games and unfolding form games. These scenarios include situations in which one or more "werewolves" are present and act as a source of fake news. The main focus was to understand how cocktail party effects and repetition dilemmas affect agent gains and how agents arrive at evolutionary stability strategies (ESS).

**Cocktail Party Effects and Repeated Dilemmas**

The cocktail party effect indicates how many other agents an agent is connected to and affects the speed and extent of information diffusion. However, our analysis showed that agents with high cocktail party effects do not necessarily have higher gains. This suggests that the quality of information may have a greater impact on gain than the quantity of information. The repetition dilemma represents the agent's choice between cooperative or noncooperative behavior, and this choice has a direct impact on the spread of fake news. However, we found no direct correlation between the value of the repetition dilemma and the gain. This suggests that agents' strategy choices depend on many factors.

**Influence of Werewolves**

In scenarios where one werewolf is present, this werewolf can manipulate the flow of information and have a noticeable impact on the overall dynamics. Understanding the impact of a werewolf's strategy on the strategy choices of other agents is important for a better understanding of strategic interactions and their evolution in the context of incomplete information games. In scenarios with multiple werewolves, agents must make action decisions under even higher uncertainty, resulting in complex patterns of information propagation, including interactions between werewolves. In such an environment, agents must employ sophisticated strategies to obtain accurate information while minimizing the risk of fake news.

### Comprehensive Considerations

Our analysis represents an important step in understanding the dynamics of information propagation and social interactions, particularly providing insights into the mechanisms of fake news diffusion. Future research is expected to validate the validity of the model and enhance its applicability in the real world by incorporating more detailed experimental data and real-world case studies. We hope that this study will contribute to the development of more effective strategies and policies to counter fake news.

To deepen the discussion of this study, we provide below the theoretical framework, the equations used, and specific examples of parameters. This will further enhance understanding of the dynamics of information propagation and social interactions.

### Theoretical Framework

Our analysis is based on the theory of incomplete information games and extensive-form games. In incomplete information games, players do not have complete information about other players' types or choices. In extensive-form games, decisions of players are made sequentially, and the choices of each player are visualized by the game tree structure.

### Equations Used

To model the influence of the cocktail party effect, we considered the connectivity $C_i$ of agent $i$. This is represented by the following equation:

$$C_i = \sum_{j \neq i} a_{ij}$$

where $a_{ij}$ equals 1 if there is a connection between agents $i$ and $j$, and 0 otherwise.

The degree of cooperation among agents in the prisoner's dilemma is modeled using the parameter $D_i$. This represents the strength of the dilemma faced by agents and is randomly assigned as follows:

$$D_i = \text{rand}(0, 1)$$

where $\text{rand}(0, 1)$ is a uniformly distributed random number between 0 and 1.

The contamination risk of fake news depends on the prisoner's dilemma and affects the payoff $P_i$ of agent $i$. This is represented by the following equation:

$$P_i = 1 \cdot D_i$$

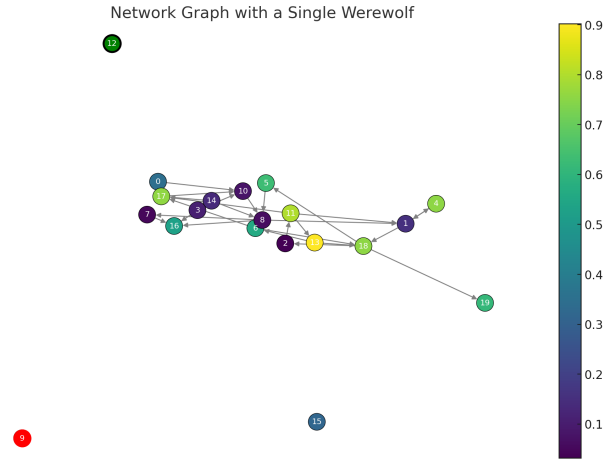

Fig. 15: Network Graph with a Single Werewolf

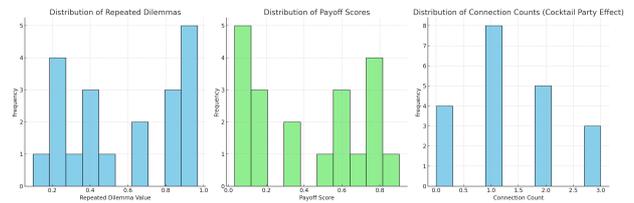

Fig. 16: Correlation between Payoff Scores and Cocktail Party Effect

where $P_i$ represents the payoff of agent $i$.

Number of agents $n = 20$ Number of wolves $w = 1$ or more The value of the prisoner's dilemma for each agent $D_i$ is assigned by a uniformly distributed random number between 0 and 1. The connectivity $C_i$ is calculated based on the number of edges between agents.

Using this theoretical framework and equations, it is possible to quantitatively analyze the complex dynamics of information propagation and social interactions and understand the strategic responses of agents to the diffusion of fake news. It is important for future research to validate these models using actual social network data and evaluate their applicability in more realistic scenarios.

The values of repeated dilemmas are distributed differently between 0 and 1. This means that agents repeatedly face dilemmas to varying degrees.

Although the payoff scores are distributed in different ranges, it appears that many agents have medium to high payoff scores. This suggests that many agents have a relatively low contamination risk.

The distribution of the number of connections shows that many agents have some connections, but it appears that few agents have very many connections. This indicates that social interactions within the network are limited.

There is no clear correlation between the value of repeated dilemmas and the number of connections. This suggests that repeated dilemmas do not have a direct impact on the agent's position within the social network.

There is also no clear correlation between gain score and number of connections. This shows that just because an agent has many connections does not necessarily mean it will get a high payoff score.

These results suggest that an agent's degree of repeated dilemmas or position in the social network do not directly affect payoff scores. An agent's payoff can be influenced not only by repeated dilemmas and social interactions, but also by other factors.

The visualized network graph shows the effects of cocktail party effects and repeated dilemmas under a non-complete information game and a non-cooperative game with one werewolf. The graph depicts the following elements: Each agent is represented as a node in the network, and the color of a node indicates the gain of that agent. The color shade is defined by the colormap 'viridis', with darker colors representing lower gains and lighter colors representing higher gains. Werewolves are represented by red nodes, distinguishing them from other agents. Werewolves have no direct interaction with other agents and play a special role in the game. The non-werewolf agent with the highest gain is represented by a green node and highlighted as an ESS. This agent is considered to have the most adapted strategy in the current environment. Edges indicate interactions between agents, with arrows indicating direction. In this non-cooperative game, agents act independently and try to maximize their gains through interactions with other agents.

Distribution of Repeated Dilemmas, The histogram shows that the values for repeated dilemmas are spread across the range but with some peaks, suggesting certain values occur more frequently than others. This could mean that some scenarios within the model are more likely to repeat or that certain interactions have a predisposition to occur.

Payoff scores are concentrated around the mid-range values, with fewer occurrences of very low and very high scores. This might indicate that the system is balanced to an extent, where extreme payoffs (both low and high) are less common. This could imply a level of equity or that the game rules are designed to moderate extreme outcomes.

Distribution of Connection Counts (Cocktail Party Effect) Most agents seem to have 2 or 3 connections, with fewer agents having only 1 or more than 3 connections. This indicates a network that is neither too sparse nor too dense, suggesting that agents have a moderate number of interactions within the system.

The network graph illustrates the connections between agents with one agent labeled as a werewolf. The color gradient likely represents payoff scores or some other measure of performance or influence. Node "9" is distinctly colored, which could indicate the werewolf in this network. This node's position and color could provide insights into its influence or success within the network. The gradient from blue to yellow across nodes might indicate varying levels of a certain metric (e.g., trustworthiness, information quality, payoff score). If it's a payoff score, nodes with a yellowish tint are achieving higher scores compared to those with a blue tint. The structure of the network is not fully connected, showing that certain nodes act as hubs or connectors to others. This could represent strategic positions within the network, where certain agents are critical for information flow or influence. The graph also shows that not all nodes are equally connected, which could impact the spread of information or misinformation. Some nodes serve as bridges and may have a more significant role in the dynamics of the network.

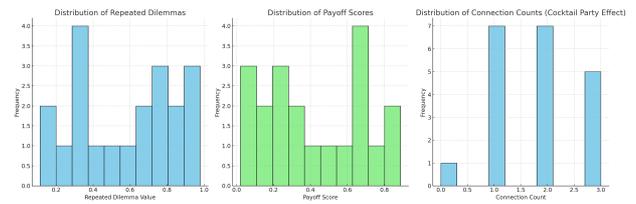

Fig. 17: Network Graph under Non-Perfect Information in Extensive-Form Game with a Single Werewolf

The payoff scores and connection counts do not seem to have a direct correlation, suggesting that other factors may be influencing success in the network. The quality of connections, the role of agents, and their strategies might be as important as the number of connections. The presence of the werewolf in the network graph does not appear to correlate with the highest payoff scores, suggesting that the role of a werewolf might not be to maximize personal payoff but possibly to influence the network in other ways, such as spreading misinformation. The network graph indicates that while some agents are well-connected and could potentially spread information effectively, the overall success or payoff is not solely dependent on these connections. It also suggests that there may be a strategy element involved in how agents interact within the network, which affects their success. The distribution of repeated dilemmas might be indicative of certain patterns or strategies within the game. Agents might be encountering the same dilemmas due to the nature of their connections or the strategies they employ.

The values of repeated dilemmas are distributed differently between 0 and 1. This means that agents repeatedly face dilemmas to varying degrees.

Although the payoff scores are distributed in different ranges, it appears that many agents have medium to high payoff scores. This suggests that many agents have a relatively low contamination risk.

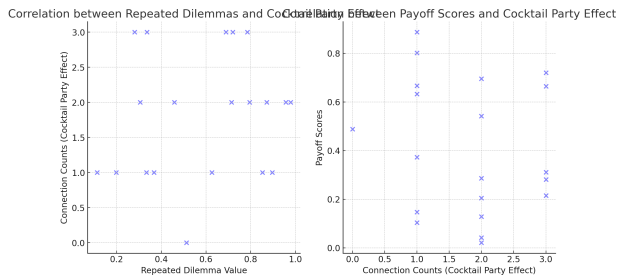

Fig. 18: Correlation between Repeated Dilemmas and Cocktail Party Effect

The distribution of the number of connections shows that many agents have some connections, but it appears that few agents have very many connections. This indicates that social interactions within the network are limited.

There is no clear correlation between the value of repeated dilemmas and the number of connections. This suggests that repeated dilemmas do not have a direct impact on the agent's position within the social network.

There is also no clear correlation between gain score and number of connections. This shows that just because an agent has many connections does not necessarily mean it will get a high payoff score.

These results suggest that an agent's degree of repeated dilemmas or position in the social network do not directly affect payoff scores. An agent's payoff can be influenced not only by repeated dilemmas and social interactions, but also by other factors.

This scatter plot appears to show data points that represent the relationship between the values associated with repeated dilemmas and the connection counts of agents. The plot does not show any clear trend or correlation, indicating that the frequency or value of repeated dilemmas does not seem to have a straightforward relationship with the number of connections (Cocktail Party Effect) that an agent has.

The second scatter plot is aiming to show the relationship between agents' payoff scores and their connection counts. Similar to the first plot, there is no obvious trend indicating a strong correlation between the two variables. This suggests that having more connections does not necessarily lead to higher payoff scores.

The histogram displays the frequency of different values associated with repeated dilemmas. It seems to have a somewhat uniform distribution with slight variations. This could imply that in the model or game being analyzed, repeated dilemmas occur with varied frequency or impact, and no single value range is predominant.

This histogram shows the distribution of payoff scores among agents. The distribution has a central concentration around the mid-range values and tails off toward the higher and lower scores. This indicates that most agents receive a payoff that is around the mid-point of the possible range, with fewer agents achieving very high or very low scores.

The last histogram shows the distribution of connection counts for agents. There is a noticeable concentration at higher connection counts, suggesting that a significant number of agents have a high number of connections within the network.

Combining insights from the scatter plots and histograms, we can infer several points, There is no clear direct correlation between the frequency or value of repeated dilemmas and the number of connections agents have, nor between the connection counts and the payoff scores. This suggests that other factors may play a more critical role in determining the outcomes or benefits that agents receive within this system. The central concentration of payoff scores indicates that while there may be variability in how agents achieve their payoffs, the system tends to reward agents in a somewhat balanced manner, with extremes being less common.

Although a large number of agents have many connections, these connections do not directly translate to payoff scores, implying that the quality or strategic use of these connections may be more important than the quantity.

The analysis of repeated dilemmas could suggest that the agents are engaged in a system where individual encounters or interactions do not have a uniform impact. This variability could be crucial for understanding the dynamics of the system as a whole.

To provide more detailed interpretations, it would be helpful to have context about the model or simulation, definitions of the metrics used (like "Repeated Dilemma Value"), and the rules that govern the interactions and payoffs within the system.